\documentclass[epj,twocolumn]{webofc}
\woctitle{The Innermost Regions of Relativistic Jets and Their Magnetic Fields}
\begin{document}
\title{Monitoring of multi-frequency polarization of gamma-ray bright AGNs}
%
%

\author{Sang-Sung Lee\inst{1}\fnsep\thanks{\email{sslee@kasi.re.kr}} \and
        Myounghee Han\inst{1} \and
        Sincheol Kang\inst{1} \and
        Jungmin Seen\inst{1}\ \and
        Do-Young Byun\inst{1} \and
        Jun-Hyun Baek\inst{1} \and
        Soon-Wook Kim\inst{1} \and
        Jeong-Sook Kim\inst{1}
}

\institute{ 
Korea Astronomy and Space Science Institute, 
Daedeokdae-ro 76, Yuseong, Daejeon 305-348,
Republic of Korea
          }

\abstract{
We started two observing programs with the Korean VLBI Network (KVN)
monitoring changes in the flux density and polarization of relativistic jets
in gamma-ray bright AGNs simultaneously at 22, 43, 86, 129~GHz.
One is a single-dish weekly-observing program in dual polarization with
KVN 21-m diameter radio telescopes beginning in 2011 May.
The other is a VLBI monthly-observing program with the three-element VLBI
network at an angular resolution range of 1.0--9.2 mas beginning in
2012 December. The monitoring observations aim to study correlation of
variability in gamma-ray with that in radio flux density and polarization
of relativistic jets when they flare up. These observations enable us
to study the origin of the gamma-ray flares of AGNs.
}
\maketitle
\section{Introduction}
\label{intro}

Many active galactic nuclei (AGNs) show strong and variable radiation
over all wavelengths. They are known to arise from a relativistic
and Doppler-boosted jet with narrow viewing angles to our
line of sight~\cite{ulr+97}.
AGNs have bipolar plasma jets relativistically originating
from supermassive black holes with an accretion disk.
Such jets are collimated by magnetic fields twisted by differential rotation
of the black hole's accretion disk or inertial-frame-dragging
ergosphere~\cite{bp82,bz77,mei+01}.
The flow velocity increases outward along the jet
in an acceleration and collimation zone containing the coiled magnetic field.
Observations of the strong and variable radiation (or outbursts)
at multi-wavelength
can potentially probe the zone~\cite{mar+08}.
The correlation between optical-to-radio outbursts and gamma-ray flares
often observed in AGNs
suggests that the high-energy emission region shall be co-spatial
with the radio knots, a few parsecs away from the black hole.
The nature of the high-energy flares can also be addressed by multi-wavelength
campaign of polarization observations on flaring AGNs.  

\section{MOGABA and iMOGABA with KVN}
\label{sec-1}

The Korean VLBI Network (KVN) is a three-element Very Long Baseline Interferometry (VLBI) network in Korea, which is dedicated to VLBI observations at millimeter wavelengths~\cite{lee+11}. Three 21-m radio telescopes are located in Seoul, Ulsan and Jeju island, Korea; KVN Yonsei Radio Telescope (KY), KVN Ulsan Radio Telescope (KU), and KVN Tanma Radio Telescope (KT) (Figure1). The baseline lengths are in a range of 305--476 km. All antennas are identical. The antennas are equipped with the quasi-optic system that allows simultaneous observations at 22, 43, 86, and 129 GHz. This system is described in detail in \cite{han+08,han+13}.

We started two observing programs of monitoring the flaring AGNs
at multi-frequencies (22, 43, 86, 129~GHz)
using the KVN.
One is a single-dish weekly-observing program in dual polarization,
beginning in 2011 May.
The observing program is called as MOGABA (MOnitoring of GAmma-ray Bright
AGNs), starting with dual-polarization observations at 22 and 43~GHz bands.
A total of $\sim$30 target sources were selected based on a sample of 
monitoring sources by the Fermi LAT space telescope.
The operating frequency of the MOGABA program has been extended
to 86 and 129~GHz bands in late 2012. 
In polarization observations, Stokes I, Q, U, V measurements were done with digital spectrometer. The total bandwidth for the observations is 512MHz.
Instrumental polarization was calibrated by observing unpolarized calibrators such Jupiter, Saturn, Mars, Venus and 3C~84 once per day. Polarization angle was referenced with measuring the polarization angle of CRAB nebula (P.A. ~ 154 deg for all frequencies) once a day. A standard polarization calibrator, 3C~286 was observed once per month for checking the calibration reliability.

The other program is, interferometric MOGABA (or iMOGABA),
a VLBI monthly-observing program with the three-element VLBI
network at an angular resolution range of 1.0--9.2 mas, beginning in
2012 December. The target sources of the iMOGABA program are
almost identical to that of MOGABA program. Full bandwidth of 256~MHz
is evenly divided into four frequency bands at 22, 43, 86, and 129~GHz
in single polarization.

The monitoring programs, MOGABA and iMOGABA, aim to study the correlation of
variability in gamma-ray with that in radio flux density and polarization
of relativistic jets when they flare up. 
The variability in gamma-ray is known to be correlated
with optical polarization angle and radio total flux for several relativistic jets~\cite{abd+10,mar+08}. Previous studies revealed that 
the gamma-ray flares are originated where the variation of optical polarization
or radio total flux happen in the relativistic jets. This implies that
the gamma-ray flares are related with changes of magnetic field in the relativistic jets.  
Since there is no dense monitoring observations of radio polarizations for the 
gamma-ray bright AGNs, we expect that the MOGABA and iMOGABA programs
will enable us to study the correlation of the gamma-ray flares with 
changes of magnetic fields in radio emission regions of the relativistic jets.

\section{Some results of the monitoring}
\label{sec-2}

\subsection{A flaring gamma-ray blazar 3C~454.3}
\label{sec-2-1}

\begin{figure}[!htbp]
\centering
\includegraphics[width=8cm,clip]{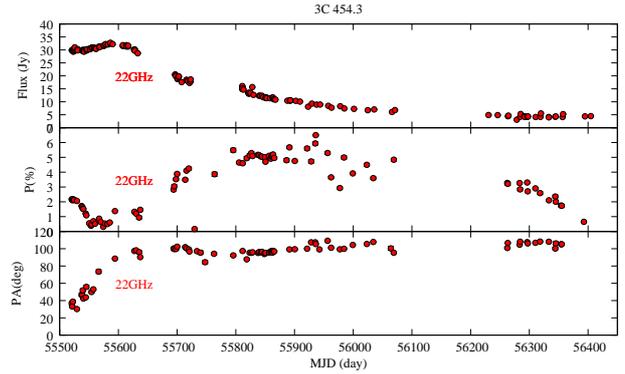}
\caption{Light curves of 3C~454.3 with KVN single-dish telescopes
at 22~GHz. Total flux (top), fractional linear polarization (middle),
and polarization angle (bottom) during MJD 55520 to MJD 56400.}
\label{fig-1}  
\end{figure}

The flat-spectrum radio quasar, 3C~454.3 (redshift $z$=0.859),
is one of most variable AGNs, showing remarkably high activity
since 2000. Recently, the unprecedented gamma-ray flares
of 2010 November 17-21 (MJD 55517 to MJD 55522) 
were detected by Fermi GST~\cite{abd+11}. 
Multi-wavelength observations have been conducted for the gamma-ray
flares~\cite{ver+11,weh+12,jor+13}.
Since MJD 55520, we also began to observe the source at 22~GHz in dual polarization
with the KVN single-dish radio telescopes,
in cadences of 3--7 days.
We found, as shown in Figure~\ref{fig-1},
that after the peak of the gamma-ray flaring (MJD 55520),
the total flux has slightly increased upto $>$30~Jy for about 100 days
and decreased gradually down to 5~Jy.
These results imply that after the peak of the gamma-ray flares the radio flux correlate
with the gamma-ray in long time scales. However,
there are time delays between the peaks of gamma-ray and radio light curves (Lee et al. in prep.).
The fractional linear polarization has decreased by a factor of 4
from 2\% to 0.5\% in about 30 days after the gamma-ray peak
and increased back with a peak of $\sim5\%$. 
Interestingly, for 100 days after the gamma-ray peak,
the polarization angle
had changed by $\sim60$~degrees, and remained almost constant for long time.
These imply that the giant gamma-ray flares may be closely related to 
events which caused the change in polarization at 22~GHz.

\begin{figure}[!htbp]
\centering
\includegraphics[width=8cm,clip]{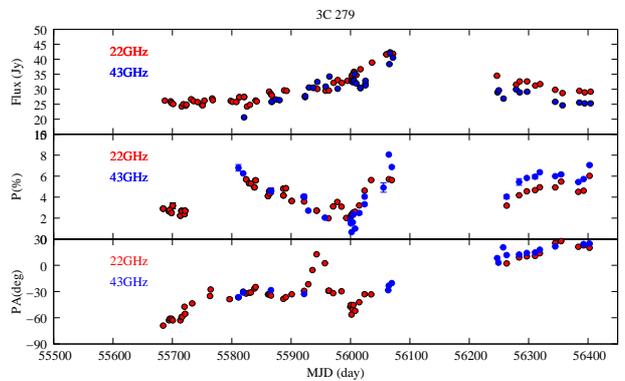}
\caption{Light curves of 3C~279 at 22~GHz (red dots) and 43~GHz (blue dots).
Total flux (top), fractional linear polarization (middle),
and polarization angle (bottom) during MJD 55520 to MJD 56400.}
\label{fig-2}  
\end{figure}

\subsection{3C~279}

One of the brightest radio sources, 3C~279 (redshift $z$=0.536),
is well known for its variations over the entire electromagnetic
spectrum with timescales from hours to years~\cite{lar+08,
col+10,hay+12}.
Abdo et al.~\cite{abd+10} discovered
the dramatic change of the optical polarization
coincident with gamma-ray flare,
providing evidence for co-spatialty of optical and
gamma-ray emission regions and indicating a highly
ordered jet magnetic field.
The source has been monitored in MOGABA program since 2011 May.
In Figure~\ref{fig-2}, we show the light curves
of 3C~279 at 22~GHz and 43~GHz bands.
The total fluxes has gradually increased from $\sim$25~Jy to $\sim$40~Jy.
The maximum of the light curves do not represent their peaks
due to the maintenance period of KVN. 
However, we found that the spectral index between 22~GHz and 43~GHz bands
changed after the maintenance period.
The fractional linear polarizations show two maximum and one minimum.
Again, we are not in a position to determine the peak of the light curves
in the fractional linear polarizations, but the minimum can be considered
as a global minimum (MJD 56000).
An interesting behavior is shown in the polarization angle at 22~GHz.
The behavior is a swing-like change of polarization angle
by $\sim$42 deg during MJD 55922 to MJD 55962.  
The polarization angle at 22~GHz has changed from -30 deg to +12 deg during the first 20 days and changed back to -30 deg during the last 20 days. 
There are no strong gamma-ray flares in the period.

\subsection{BL Lacertae}

\begin{figure}[!t]
\centering
\includegraphics[width=8cm,clip]{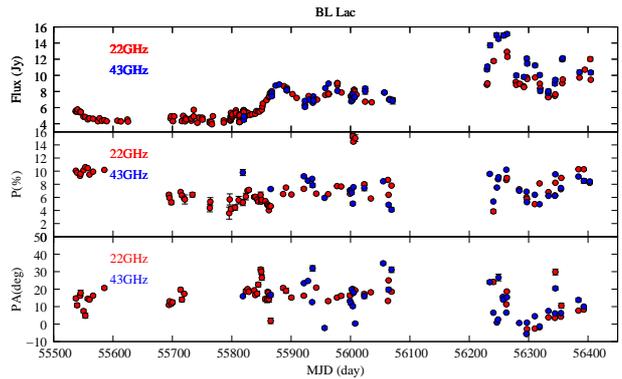}
\caption{Light curves of BL Lac at 22~GHz (red dots) and 43~GHz (blue dots).
Total flux (top), fractional linear polarization (middle),
and polarization angle (bottom) during MJD 55520 to MJD 56400.}
\label{fig-3}  
\end{figure}

BL Lacertae is one of the bright, highly variable AGNs (known as blazars)
at low redshift ($z$=0.069, \cite{mh77}). As all blazars, it shows
strong flux and spectral variability at all wavelengths and on a variety of time scales.
BL Lacertae started a series of strong flares
in 2011 May after moderately active phase at gamma-rays and
optical frequencies (see \cite{arl+13},\cite{rai+13} and references therein).
It underwent a very rapid (a decaying time of 13~min) TeV gamma-ray flaring
on MJD 55740 (2011 June 28)~\cite{arl+13}.
We began to monitor the object in 2010 November. 
The total flux, as shown in Figure~\ref{fig-3}, started to increase
in MJD 55800, peaking in $\sim$MJD 55880
both at 22~GHz and 43~GHz. The spectral index between two frequencies
has been quite flat through the increase. 
The peak is coincident with one of gamma-ray flares.
We found also that the source already entered a very active state since MJD 56200
after a long maintenance of KVN.
The fractional linear polarization has been decreased from 10\% to 5\%
between MJD 55600 and MJD 55700. 
Interestingly, during MJD 56003 to MJD 56007, there was a abrupt increase of 
the fractional linear polarization by a factor of 2 (from 7\% to 15\%) at 22~GHz.
However, we do not see such a change at 43~GHz light curve.

\subsection{OJ 287}

\begin{figure}[!t]
\centering
\includegraphics[width=8cm,clip]{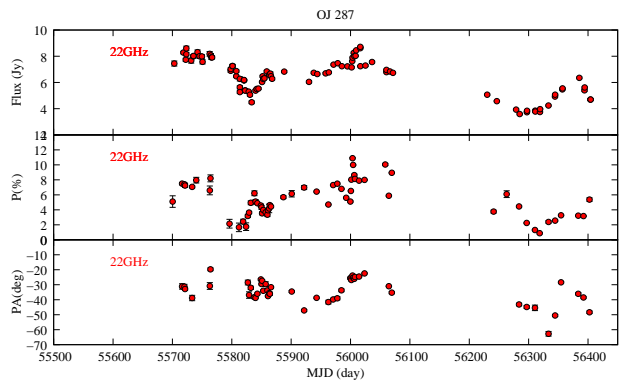}
\caption{Light curves of OJ~287 with KVN single-dish telescopes
at 22~GHz. Total flux (top), fractional linear polarization (middle),
and polarization angle (bottom) during MJD 55520 to MJD 56400.}
\label{fig-4}  
\end{figure}

OJ~287 ($z$=306) is one of the well-studied BL Lacertae objects,
known to produce pairs of optical outbursts every 12 years~\cite{sil+88}.
Many studies interpreted this system as a supermassive black hole (SMBH)
binary~\cite{sil+88,lv96,val+06,val+08}.
Agudo et al.~\cite{agu+11} found a strong correlation between 
gamma-ray and the millimeter emission during two major gamma-ray flares
in 2008 and 2009, concluding the gamma-ray emission region
must be at a distance $>$14 pc from the central engine.
We began to monitor OJ~287 in MOGABA program since 2011 May (MJD 55700). 
During MJD 55800 to MJD 55850, an apparent anti-correlation between fractional linear polarization and total flux density at 22~GHz was observed as shown in Figure~\ref{fig-4}. 
We note that on MJD 55845 there was gamma-ray flares detected by the Large Area Telescope (LAT) on the Fermi
Gamma-ray Space Telescope (ATel. \#3680).
This implies that the anti-correlation between the total flux and fractional
linear polarization may be closely related with the gamma-ray flares,
implying that the anti-correlation could be considered as a pre-cursor
of the gamma-ray flares.

\subsection{Polarization Measurements in Microquasar Cygnus X-3}

Since 2013 January, we have monitored a microquasar, Cygnus X-3
in order to study polarization characteristic in X-ray binaries when flaring up.
There was one observation with a marginal detection for fractional linear polarization
of $\sim1.3\%$ when Cygnus X-3 was a few times brighter
than typical radio flux density of ~100 mJy or less.
A new giant flare would be expected in late 2013 or in 2014 with the consideration
of the long-term multi-wavelength monitoring in Cygnus X-3.
Therefore, a higher degree of polarization is expected during the upcoming
giant flare, and we are preparing a Target-of-Opportunity observation.

\section{Summary}

In order to address the nature of high-energy flares in AGNs, we started to run two observing programs (MOGABA and iMOGABA) for monitoring changes of the flaring gamma-ray AGNs in linear polarization and flux density at 22/43/86/129 GHz at angular scales of arcsecond-to-milliarcsecond Korean VLBI Network. We detected interesting changes in the fractional linear polarization and polarization angle
of 3C~454.3, 3C~279, BL Lac and OJ~287. Some of the changes may correlate 
with the gamma-ray flares. We expect that the on-going multi-frequency polarization monitoring programs with KVN will provide us with important information of 
magnetic field characteristics of the relativistic jets when they flare up in high energy. 
\\\\\\
{\bf Acknowledgments}\\

\noindent
We are grateful to all staff members in KVN who helped to operate the array and to correlate the data for MOGABA and iMOGABA programs. 
The KVN is a facility operated by the Korea Astronomy and Space Science Institute.
This work was supported by Global Research Collaboration and 
Networking program of Korea Research Council of Fundamental Science \& Technology (KRCF).

\end{document}